\documentclass[conference]{IEEEtran}
\usepackage{blindtext}
\usepackage{graphicx}
\usepackage{color}
\usepackage{gensymb}
\ifCLASSINFOpdf
\else
\fi
\begin{document}
\title{A Lightweight Blockchain-based Privacy Protection for Smart Surveillance at the Edge}

\author{
\IEEEauthorblockN{Alem Fitwi${^\dagger}$, Yu Chen${^\dagger}$, Sencun Zhu${^\ddagger}$}
\IEEEauthorblockA{${^\dagger}$Dept. of Electrical \& Computer Engineering,
Binghamton University, SUNY,  Binghamton, NY 13902, USA \\
${^\ddagger}$Dept. of Computer Science and Engineering, Penn State University, University Park, PA 16802, USA\\ 
Emails: \{afitwi1, ychen\}@binghamton.edu; szhu@cse.psu.edu}
}
\maketitle
\begin{abstract} 

Witnessing the increasingly pervasive deployment of security video surveillance systems(VSS), more and more individuals have become concerned with the issues of privacy violations. While the majority of the public have a favorable view of surveillance in terms of crime deterrence, individuals do not accept the invasive monitoring of their private life. To date, however, there is not a lightweight and secure privacy-preserving solution for video surveillance systems. The recent success of blockchain (BC) technologies and their applications in the Internet of Things (IoT) shed a light on this challenging issue. In this paper, we propose a Lightweight, Blockchain-based Privacy protection (Lib-Pri) scheme for surveillance cameras at the edge. It enables the VSS to perform surveillance without compromising the privacy of people captured in the videos. The Lib-Pri system transforms the deployed VSS into a system that functions as a federated blockchain network capable of carrying out integrity checking, blurring keys management, feature sharing, and video access sanctioning. The policy-based enforcement of privacy measures is carried out at the edge devices for real-time video analytics without cluttering the network. 

\end{abstract}

\IEEEpeerreviewmaketitle

\section{Introduction}
\label{sec:intro}


Video surveillance cameras are pervasive in public places in response to growing security concerns \cite{nikouei2018real, xu2018real}. According to a report produced by IHS Markit, there are about 245 millions surveillance cameras in operation today. London is the city with the highest number of cameras, where an average Londoner is estimated to be caught on camera 300 times a day. The number in China is expected to grow more than three times by the year 2020 \cite{cavallaro2007privacy, wang2018enabling}. Clearly individuals' privacy is at stake! People are being observed with or without their awareness almost wherever they go. This situation widely incurs concerns in the violation of individual's privacy.

The more powerful the modern surveillance cameras become, the more likely they are to be abused to gather private information.  Authorized security personnel in charge of the surveillance system might abuse the cameras for voyeurism, cyber stalking, and unauthorized collection of data on activities or behaviors of individuals \cite{streiffer2017eprivateeye, wang2018enabling, yu2017iprivacy}. Maneuverable cameras, like pan-tilt-zoom (PTZ) camera, could be abused and directed to intrusively spy on other people in their apartments. For instance, there was an investigation launched after a security guard spied on the private apartment of the German Chancellor Angela Merkel using a museum's closed circuit television (CCTV) camera  \cite{cavallaro2007privacy, hessler2006museum}. Obviously, while the benefits of surveillance greatly outweigh the potential risks, surveillance and the privacy of people should be balanced out. There has been a number of efforts to address the privacy requirements through the introduction of smart cameras with embedded privacy curtailments in lieu of trying to abandon the practice of surveillance \cite{dufaux2011video, streiffer2017eprivateeye, wang2018enabling, yu2017iprivacy, zhang2013chaos}. However, a resource and bandwidth aware privacy-protection mechanism is still missing in most surveillance camera systems today. 

The Bitcoin has substantiated how the BC technology can decentralize trusted computing models \cite{nakamoto2008bitcoin}. A light-weight blockchain along with the concept of identity-based distributed data possession in multi-cloud storage can be leveraged to ensure privacy and authorized access in smart surveillance \cite{dorri2017lsb, kshetri2017blockchain, nagothu2018microservice, wang2015identity}. Due to concerns on privacy and performance issues, however, public blockchain is not an ideal candidate. Instead, a private blockchain has been considered where only authenticated member nodes join \cite{do2017blockchain, zyskind2015decentralizing}. Therefore, a lightweight, closed-group blockchain that supports decentralized applications like surveillance that entails high speed and privacy could meet the requirements. 

In this position paper, a lightweight blockchain-based privacy protection (Lib-Pri) scheme for smart surveillance at the edge is proposed. It enables the construction of a privacy-aware smart surveillance system by integrating the advanced features of BC and smart contract with object detection (OD) technologies coupled with image scrambling techniques. 
The Lib-Pri system consists of three major parts in this privacy and resource aware service: smart cameras, BC nodes and users. The smart cameras are the edge devices with embedded configurable set of privacy policies. The cameras capture videos and process them using the attached single board computer (SBC), e.g. the new Jetson Nano Module, a Tinker board or Raspberry Pi. Privacy-sensitive objects are detected and corresponding privacy protection measures are enforced. The BC network ensures authenticity. Users are assigned with different levels of access privileges to the videos, which are defined in the smart contracts. 
In addition, an isolated storage system is considered for the storage of reversibly scrambled images or videos for law enforcement purposes. 

The remainder of this paper is structured as ensues. The architecture of the Lib-Pri system is described in Section \ref{sec:bchain} followed by the explication of the privacy polices in Section \ref{sec:policy}. At last, the conclusive remarks and the direction of our ongoing research are presented in Section \ref{sec:con}.

\section{Lib-Pri system: Architecture and Components}
\label{sec:bchain}

\begin{figure}[t]
    \centering
        \includegraphics[width=0.425\textwidth]{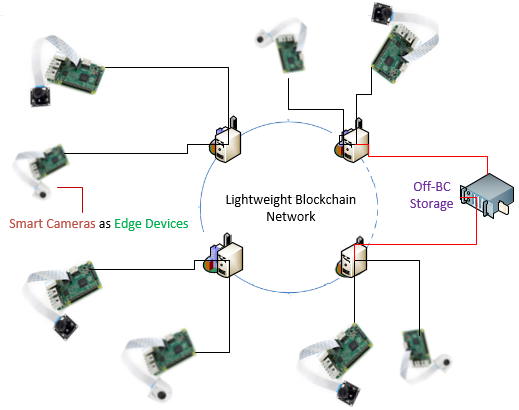}
    \caption{Privacy-preserving Lite Blockchain-based Smart Surveillance.}
    \label{fig_1}
    \vspace{-10pt}
\end{figure}

Figure \ref{fig_1} portrays the high-level overview of the proposed Lib-Pri system, the Lightweight BC-based Privacy Protection for smart surveillance at the edge. It comprises smart cameras as edge devices, BC nodes, authorized users such as security personnel or law enforcement officials, and off-BC storage. 

\subsection{Smart Surveillance Cameras At The Edge}
\label{sec:SSCE}

The smart cameras function as edge devices. They capture videos, perform frame splitting and reversible blurring following an object detection as per the defined policies configured in each camera. A pre-trained machine learning (ML) neural network (NN) is loaded onto the edge for multiple objects detection within a budgeted time. Features are extracted for OD, detected objects that match those on the policy are bounded, a blurring key is generated and securely shared by means of a public crypto-system, and then the sensitive objects are chaotically masked at the edge. The smart cameras also support the opt-out policy when no attention is paid to privacy. This boosts the public confidence that their images or videos captured by the surveillance cameras will not be abused. The opt-out option is defined in the smart contract, and it renders the privacy policy useless when triggered by authorized users. 

Many of the existing CCTV systems are centralized and not ``smart''. Videos analytics are performed at a centralized locations. With the mounted SBCs, it is feasible to deploy and run pre-trained object detection ML algorithms on edge devices. Performing video analytics, information distillation, object detection, and human face denaturing at the edge drastically reduces the cameras egress bandwidth consumption. Meanwhile, edge devices still suffer from resource constraints. As a consequence, at times when the smart cameras fail to meet the requirements of real-time video analytics, adopting a hybrid of edge and fog computing paradigms is an alternative. Some of the tasks will be offloaded to a near-site fog node, i.e. a laptop, a tablet, or a smartphone. 

\subsection{The Lightweight Blockchain Network}
\label{sec:LBC}

A permissioned, private BC network is envisioned in the Lib-Pri system, which is not open to public access. 
The Lib-Pri system employs privacy-preserving smart contracts that define privileges and access rules, which allow authorized users access the videos without violating the privacy of individuals. 

The existing smart contract however doesn't preserve the privacy and confidentiality of personal files or data when shared to the BC nodes. In our proposal, the smart contract is tailored for better handling of the privacy and confidentiality issues. Users who try to access the video need to prove themselves to the blockchain that they have the permission. Besides, correctness of the references are validated by every other node so that only authorized accesses are made and the integrity of the video storage is verifiable. The reference in this paper points to a variable on a mapping table that contains actual references to the videos.

\subsection{Off-Blockchain Storage}
\label{sec:SOS}
Storing data on the blockchain is expensive and the storage requirements are impractical if video streams are stored in replicated forms in every node of the blockchain network. As a solution, a storage server/servers outside the blockchain network is considered that could be connected to the blockchain. The connection is made as per the business rules defined on the smart contract. Only videos and images deemed important for later law enforcement purposes are stored in the off-network storage. Video frames containing objects that show offensive behaviors and fugitives are marked according to the predefined policies and are pushed to the storage. The hashed references and germane access details are stored on the blockchain. Besides, access histories important for auditing are logged and stored on the off-BC storage. Only their references are similarly shared to the BC nodes. 

The blurring keys sharing is managed using a public key crypto-system in two stages; between smart cameras and BC nodes, and between the BC nodes and the off-BC storages as necessary. When there is no need to store videos, the first stage of key management suffice. The second stage is required only when videos have to be stored for later use. Then, keys are revoked the moment the corresponding video is deemed unwanted and deleted. 

\subsection{Object Detection, Bounding And Masking}
\label{sec:ODB}
Mechanisms that could perform live video frame denaturing for privacy purpose are required to be robust and fast. Therefore, the accuracy, sophistication, and speed of live-video processing method is vital. The functionality of our Lib-Pri system includes detecting objects in an image or video frame, adding bounding box, and applying reversible chaotic mask to preserve privacy. 

Object detection is a compute-intensive process. Hence, we will design the state-of-the art object detection model, and make an apropos offline training on sufficiently large data set. The trained model will be loaded to the edge devices. In order to avoid any possible violation of the budgeted time for live video analytics, the fog server works along with an edge device that supports pipe-lining and parallel processing of frames to expedite the process. 

Following the successful detection of an object and extraction of coordinates of the bounding box, masking is done using a chaotic method to conserve privacy. The method is inspired by the earlier reported researches \cite{fitwi2011performance, rahman2010real, zhang2013chaos}. The initial conditions of the chaotic generator that comprises both floating and integral values are used as the key. 

\subsection{Spotting Fugitives}
\label{sec:SF}
Our Lib-Pri system is able to identify fugitives after the first time they were caught doing something deemed an aggressive or a criminal act. Once some individuals were detected (with visible face) doing something wrong and disappeared, the authorized law enforcement people are provided with the option to flag the germane video frames or images containing misconduct as wanted. The fugitive's facial features required for comparison are dispatched to the edge with their original references pushed to the blockchain network nodes. When a match is obtained by any of the cameras demanded to spot the escapees, they will send an immediate alert message comprising the location and identity of the fugitive.
\section{Privacy Policies}
\label{sec:policy}

As far as video surveillance is concerned, a number of arguments and views have been raised pro and against of it. Covert government surveillance is one of the push factors that caused the public to grow suspicious about it. Governments often quote the famous Margalit's saying: ``\emph{If you have got nothing to hide, you have got nothing to worry about}.'' However, many argue against that it is never all about hiding something, it is all about that being none of other people's business. They argue in favor of the quote "\emph{I don't have anything to hide, but I don't have anything I feel like showing to you, either}" \cite{kumar2015promoting}. Putting it another, many people concur on the usefulness of surveillance system for security purposes but they firmly believe that the indiscriminately pervasive and expansive deployment has endangered constitutionally enshrined values like the right of anonymity, a form of privacy. People going to bars, infertility clinic, doctors, etc. want to be anonymous. But cameras in public places enable officials to non-selectively observe these routine activities of individuals irrespective of whether they are law-abiding citizens or criminals. Besides, there is a risk that recorded footage can be abused or leaked. Hence, the public want the surveillance system to be aware of privacy-sensitive things. 

Therefore, privacy and access policies are included in the Lib-Pri system to guard against the abuses of surveillance videos by authorized officials or other users, including unwarranted monitoring, peeping, and leaking. Privacy policies that specify which objects or video frames should be denatured are incorporated in smart cameras, or more precisely speaking, on the mounted edge computing devices. The privacy policies recognize the scenarios, objects, activities, or simply locations that are not permitted to be posted publicly. The fine-grained access privilege policies define who accesses what and how. 

\begin{figure}[t]
    \centering
        \includegraphics[width=0.425\textwidth]{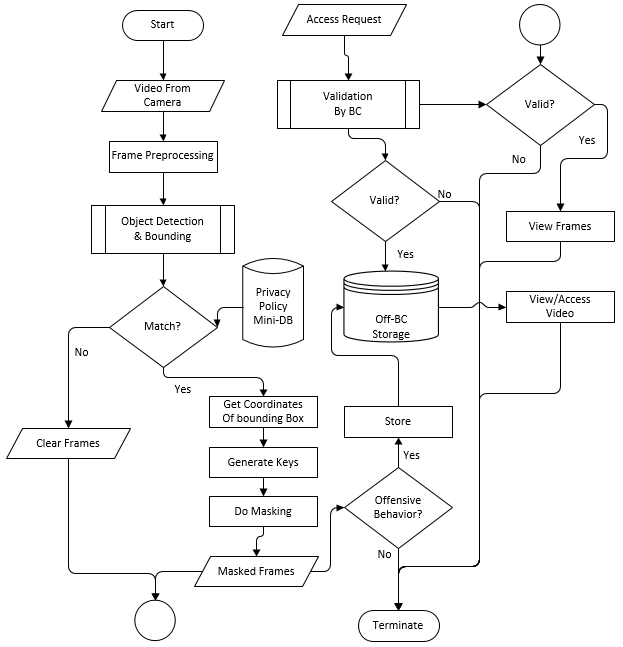}
    \caption{The Object Detection, Masking, and Access policing Processes.}
    \label{fig_2}
    \vspace{-10pt}
\end{figure}

Access control is enforced by the smart contract embedded in the BC network. As a private BC network, the access is not anonymous as what the public blockchain network allows for Bitcoin. The user has to be verified as one of the privileged users who are authorized to watch the requested videos. The policy is set to ensure that things are effectively obscured if they have the potential of revealing information pertinent to an individual's privacy. The contents or objects considered in the privacy policy of our proposed system are listed as ensues. The entire policing process is depicted in Fig. \ref{fig_2}.

\begin{enumerate}
    \item By default, the face of every captured individual is reversibly scrambled. Faces can only be recovered when proof to verify the identity of someone engaged in misdeeds is required (for example at a court of law). 
    
    \item To prevent peeping at people doing private things at home or capturing videos via windows by maneuvering cameras, objects on frames identified as windows are reversibly masked at the edge. This is an act of voyeurism unless cameras are warranted to do so in an opt-out mode. Another case where the opt-out option could be enforced is when, for instance, drones with smart cameras are on exigent or search-and-rescue mission where clear detection of humans or objects is of paramount importance and privacy is the least concern.
    
    \item Bedrooms, bathrooms, plate numbers, personal information tags, and other things that could reveal information that leads to potential privacy breach are masked.
    
    \item Fine-grained access control policies and privileges are set as part of the privacy policies. They can be legitimately and authorizedly updated or revoked at any time.
    
    \item In a frame containing nude images of individuals the face, genitals, buttocks, and breasts are scrambled.
    
    \item Video frames with aggressive behaviors like throwing a fist, pointing guns, and an act of vandalizing properties are flagged and stored for later law enforcement uses. 
    
    \item References to the identities of viewers or those who access the videos are embedded into the video frames they viewed or accessed to reduce unauthorized leaking. 
    
    \item To prevent leaking of videos or images, videos or images stored cannot be transferred or copied. They can only be viewed by authorized law enforcement people with the right privileges. Only when need arises, warranted people could do the transfer or copying activities.
    
    \item Keys generated for denaturing frames at the camera end are automatically added to the authorized viewers' details on the blockchain based on their privileges and access nuances defined on the privacy policies. A key is kept as long as a reference to the corresponding stored video exists in the BC nodes. 
    
    \item There is an opt-out option which could be triggered by authorized users whenever privacy is deemed unimportant. For instance, for live examination proctoring, the blurring option might not be required.
    
    \item Different nodes of the blockchain network could have different sets of policies. Even cameras could be set individually or zone-wise to have different privacy policies.
    
    \item The smart contracts contain attributes common across the surveillance facility on top of the attributes and requirements unique to each organization or community who own the system.
    
    \item Access histories of stored or viewed videos are logged and shared onto the blockchain based on the details defined in smart contracts. Viewer's identity, time of view or access, privileges, and references of accessed parts are logged and stored. 
\end{enumerate}

\section{Conclusions and Works in Progress}
\label{sec:con}

It is a violation of the privacy rights of individuals and a debacle of the rule of law to covertly and non-selectively monitor people regardless of whether they are suspects or not. This is how most of today's safety/security surveillance systems operate. They give the least concern to privacy.

In this position paper, we are intending to integrate and leverage the advantages of the blockchain, smart contract, object detection, and edge computing technologies to make the existing video surveillance systems privacy-aware. In the proposed Lib-Pri system, frame splitting, object detection and bounding, and policy-based enforcement of privacy measures are performed at the edge devices, i.e. the smart cameras, to enable real-time video analytics without congesting the communication network. Offensive behaviors of individuals are detected by an object-detection network on the edge device and is then offloaded to an off-blockchain storage. Fugitives are also identified by comparing previous facial features and those extracted from the live video frames. To discourage leaking of videos and images, information specific to the viewers is embedded into the videos/images automatically and a corresponding log-reference is pushed to the BC. 

Our on-going efforts consist of multiple tasks, including: 

\begin{itemize}
    \item Data set collection and labeling for the NN training;
    \item Polishing and encoding the proposed privacy policies into smart contracts; and
    \item A prototype of the Lib-Pri system is being built, which includes multiple surveillance cameras with SBCs like mounted Raspberry Pi, smartphones and laptops that function as fog devices.
\end{itemize}

\ifCLASSOPTIONcaptionsoff
  \newpage
\fi
\bibliographystyle{IEEEtranS}
\bibliography{Ref.bib}
\end{document}